\documentclass[12pt]{article}
\usepackage{epsfig,cite}
\usepackage {amsmath}
\usepackage{amssymb}
\include{epsf}
\newcount\timecount
\newcount\hours \newcount\minutes  \newcount\temp \newcount\pmhours
\hours = \time
\divide\hours by 60
\temp = \hours
\multiply\temp by 60
\minutes = \time
\advance\minutes by -\temp
\def\hour{\the\hours}
\def\minute{\ifnum\minutes<10 0\the\minutes
            \else\the\minutes\fi}
\def\clock{
\ifnum\hours=0 12:\minute\ AM
\else\ifnum\hours<12 \hour:\minute\ AM
      \else\ifnum\hours=12 12:\minute\ PM
            \else\ifnum\hours>12
                 \pmhours=\hours
                 \advance\pmhours by -12
                 \the\pmhours:\minute\ PM
                 \fi
            \fi
      \fi
\fi
}

\def\monthname{\relax\ifcase\month 0/\or January\or February\or
   March\or April\or May\or June\or July\or August\or September\or
   October\or November\or December\else\number\month/\fi}

\def\bold#1{\setbox0=\hbox{$#1$}%
     \kern-.025em\copy0\kern-\wd0
     \kern.05em\copy0\kern-\wd0
     \kern-.025em\raise.0433em\box0 }

\def\beq{\begin{equation}}
\def\eeq{\end{equation}}

\def\ga{\mathrel{\raise.3ex\hbox{$>$\kern-.75em\lower1ex\hbox{$\sim$}}}}
\def\la{\mathrel{\raise.3ex\hbox{$<$\kern-.75em\lower1ex\hbox{$\sim$}}}}
\def\gev{{\rm \, Ge\kern-0.125em V}}
\def\tev{{\rm \, Te\kern-0.125em V}}
\def\gyr{{\rm \, G\kern-0.125em yr}}

\def\gappeq{\mathrel{\rlap {\raise.5ex\hbox{$>$}}
{\lower.5ex\hbox{$\sim$}}}}
\def\lappeq{\mathrel{\rlap{\raise.5ex\hbox{$<$}}
{\lower.5ex\hbox{$\sim$}}}}
\def\Toprel#1\over#2{\mathrel{\mathop{#2}\limits^{#1}}}

\def\m12{m_{1\!/2}}

\def\NP{{Nucl. Phys.} }

\def\tanb{\tan \beta}

\def\swsq{\sin^2 \theta_W}

\def\bea{\begin{eqnarray}}
\def\eea{\end{eqnarray}}
\newcommand{\goto}{\rightarrow}
\newcommand{\bmm}{B_s \goto \mu^+ \, \mu^-}
\newcommand{\mbs}{M_{B_s}}
\newcommand{\fbs}{f_{B_s}}

\begin{document}
\begin{titlepage}
\pagestyle{empty}
\baselineskip=21pt
\rightline{\tt hep-ph/0603136}
\rightline{CERN-PH-TH/2006-037}
\rightline{UMN--TH--2434/06}
\rightline{FTPI--MINN--06/06}
\vskip 0.2in
\begin{center}
{\large{\bf On {\boldmath $B_s \to \mu^+ \mu^-$} and Cold Dark Matter Scattering
in the MSSM with Non-Universal Higgs Masses}}
\end{center}
\begin{center}
\vskip 0.2in
{\bf John~Ellis}$^1$, {\bf Keith~A.~Olive}$^{2}$,
{\bf Yudi Santoso}$^{3}$
and {\bf Vassilis~C.~Spanos}$^{2}$
\vskip 0.1in

{\it
$^1${TH Division, CERN, Geneva, Switzerland}\\
$^2${William I. Fine Theoretical Physics Institute, \\
University of Minnesota, Minneapolis, MN 55455, USA}\\
$^3${Department of Physics and Astronomy, University of Victoria,\\
 Victoria, BC, V8P 1A1, Canada}}

\vskip 0.2in
{\bf Abstract}
\end{center}
\baselineskip=18pt \noindent
%%%%%%%%%%%%%%%%%%%%%%%%%%%%%%%%%%%%%%%%%%%%%%%%%%%%%%%%%%%%%%%%%%%%%

We show that present experimental constraints on $\bmm$ decay 
and the CDMS upper limit on the cold dark matter elastic scattering
cross section already have significant impact on the parameter
space of the minimal supersymmetric extension of the Standard Model 
(MSSM) with non-universal supersymmetry-breaking scalar masses for
the Higgs multiplets (NUHM). The relaxation of scalar universality
in the MSSM allows the possibility of a relatively light mass $m_A$
for the pseudoscalar Higgs boson. The present upper limit on $\bmm$
already excludes much of the scope for this possibility in the NUHM,
in contrast to the constrained MSSM with universal scalar masses
(CMSSM), where $\bmm$ decay does not exclude any ranges of
parameters not already excluded by $b \to s \gamma$ decay. Cold
dark matter scattering is also enhanced for small $M_A$, but the impact 
of present upper limit on $\bmm$ on the NUHM parameter space is 
in many cases greater than that of the CDMS scattering limit,
particularly at large $\tan \beta$.

%%%%%%%%%%%%%%%%%%%%%%%%%%%%%%%%%%%%%%%%%%%%%%%%%%%%%%%%%%%%%%%%%%%%%
\vfill
\leftline{CERN-PH-TH/2006-037}
\leftline{March 2006}
\end{titlepage}
\baselineskip=18pt
%%%%%%%%%%%%%%%%%%%%%%%%%%%%%%%%%%%%%%%%%%%%%%%%%%%%%%%%%%%%%%%%%%%%%

\section{Introduction}

Many phenomenological analyses of the parameter space of the MSSM
assume universality for the soft supersymmetry-breaking scalar and gaugino masses,
a theoretical framework often termed the constrained MSSM (CMSSM). However,
this universality assumption is not necessarily supported by the
effective supergravity models derived, for example, from string theory.
On the other hand, the phenomenological suppression of flavour-changing
neutral interactions suggests that squarks and sleptons with the same
internal quantum numbers must be very nearly degenerate, at least for
the supersymmetric partners of the first two generations, and there
would be degeneracy (before renormalization) between squarks and sleptons
in common GUT multiplets. However, there is no strong reason to
suppose that the soft supersymmetry-breaking scalar masses 
$m_i^2, i = 1,2$ of the Higgs
multiplets should necessarily be the same as each other or the squarks
and sleptons: $m_i^2 = (1 + \delta_i) m_0^2$ with $\delta_{1,2} \ne 0$. 
These considerations motivate the phenomenological study
of models with non-universal Higgs masses
(NUHM)~\cite{nonu,oldnuhm,EFGOS,nuhm,DDnuhm,othernuhm}, as
considered in this paper.

The parameter space of the NUHM has two dimensions more than the CMSSM
that are spanned by $\delta_{1,2}$,
allowing the Higgs supermultiplet mixing parameter $\mu$ and the
pseudoscalar Higgs mass $m_A$ to be treated as parameters that are free,
apart from theoretical constraints such as vacuum stability up to the scale of
grand unification: to this end, we
impose the requirement that $m_i^2 + \mu^2 > 0$ at all renormalization
scales below this GUT scale~\cite{nuhm}.
The phenomenological constraints on the NUHM provided by LEP constraints
on the masses of the lightest supersymmetric Higgs boson $m_h$~\cite{mh} 
and the lighter chargino $\chi^\pm$ have been considered, along with
$b \to s \gamma$ decay \cite{bsg,bsgth}, the relic dark matter density
$\Omega_{CDM} h^2$~\cite{WMAP} and (optionally) the anomalous magnetic 
moment of the muon, $g_\mu - 2$ \cite{g-2}. These constraints allow regions of
NUHM parameter space in which $m_A$ is considerably smaller than its
value in the CMSSM.

The decay $\bmm$ is known to impose another interesting
constraint on the parameter spaces of models for physics beyond
the Standard Model, such as the MSSM~\cite{Dedes,Arnowitt,ko,baer}. 
The Fermilab Tevatron collider already has an interesting upper limit 
$\sim 2 \times 10^{-7}$ on the $\bmm$ decay branching ratio~\cite{cdf}, and
future runs of the Fermilab Tevatron collider and the LHC are expected to 
increase significantly the experimental sensitivity to $\bmm$ decay.
However, a recent exploration of the present $\bmm$ constraint in the
CMSSM~\cite{Bmumu} found that its impact was limited by uncertainties in the
theoretical relation of $m_A$ to the underlying CMSSM parameters, and
provided no extensions of the $(m_{1/2}, m_0)$ regions already excluded by $b \to s 
\gamma$ decay, in particular.

We explore in this paper the current impact of this additional constraint on
the NUHM, and consider also the
potential impact of future improvements in the experimental sensitivity 
to $\bmm$ decay within the NUHM. The rate for $\bmm$ may be enhanced
in portions of the NUHM parameter space where $M_A$ is smaller than in the CMSSM.
We find that, consequently, significant regions of the 
NUHM parameter space at small $m_A$ and large $\tan \beta$
are already excluded by the present
experimental upper limit on $\bmm$ decay. Likely improvements in
sensitivity at the Fermilab Tevatron collider and the LHC will reach
significant extra swathes of the NUHM parameter space.

The elastic cold dark matter scattering cross section is also enhanced
at small $M_A$, and another important constraint on the NUHM parameter 
space is placed by
the upper limit on the spin-independent cold dark matter scattering cross
section from the CDMS Collaboration~\cite{CDMS}. This has only just begun to cut 
into the CMSSM parameter
space \cite{DDupdate}, but does impact the NUHM parameter space, as also discussed in
this paper. However, in many of the specific cases studied, the $\bmm$
constraint is stronger than the CDMS constraint.

The structure of this paper is as follows. In Section 2 we recall briefly
essential aspects of the theoretical calculation of $\bmm$ decay and 
spin-independent dark matter scattering. Then, in
Section 3 we explore various slices through the NUHM parameter space,
displaying the impact of the present experimental upper limit on $\bmm$
decay. Finally, in Section 4 we discuss the potential impact of future
improvements in the experimental sensitivity to this decay.

\section{Review of the Calculations of {\boldmath $\bmm$} Decay and
Spin-Independent Elastic {\boldmath $\chi$} Scattering}

The branching ratio for the decay $\bmm$ is given by
\bea
\mathcal{B}(\bmm) &=& \frac{G_F^2 \alpha^2}{16 \pi^3} \frac{\mbs^5 \fbs^2  
\tau_B }{4} |V_{tb}V_{ts}^*|^2 \sqrt{1-\frac{4 m_\mu^2}{\mbs^2}} \nonumber \\
  &\times& \left\{    \left(1-\frac{4 m_\mu^2}{\mbs^2}\right) | C_S |^2 
  + \left |C_P-2 \, C_A \frac{m_\mu}{\mbs^2} \right |^2   \right\} \, ,
\label{eq:braratio}
\eea
where the one-loop corrected Wilson coefficients $C_{S,P}$ are taken 
from~\cite{Bobeth} and $C_A$ is defined in terms of $Y(x_t)$, 
following~\cite{Logan}, as $C_A=Y(x_t)/\swsq$ where
\beq
Y(x_t)= 0.99 \left( \frac{m_t(m_t)}{165 \gev} \right)^{1.55} \, .
\eeq
The function $Y(x_t)$ incorporates both leading \cite{Inami} and
next-to-leading order~\cite{Buras} QCD corrections, and $m_t(m_t)$ is the
running top-quark mass in the $\overline{MS}$ scheme. We assume here that
the physical top quark mass is $m_t=172.7 \pm 2.9$~GeV~\cite{mtop172}.

The Wilson coefficients $C_{S,P}$ receive four contributions in the
context of MSSM, due to Higgs doublets~\cite{Logan}, counter-terms, box and penguin
diagram~\cite{calcs,Babu}\footnote{See~\cite{Bobeth}, where the full one-loop 
corrections have been calculated.}. We have considered all these  
one-loop corrections  as well as the dominant NLO 
QCD corrections studied in~\cite{Buras2}. In addition, we have included 
the flavour-changing gluino contribution~\cite{Tata,Bobeth2}.
The Wilson coefficients $C_{S,P}$ have been multiplied by 
$1/(1+\epsilon_b)^2$, where $\epsilon_b$ incorporates the full flavour-independent 
supersymmetric one-loop corrections to the bottom-quark Yukawa
coupling~\cite{mbcor,Carena, Pierce}, that in principle are   significant in the
large-$\tanb$ regime~\cite{Dedes,Arnowitt}.
Furthermore, it is known that the flavour-violating 
contributions arising from the Higgs and chargino couplings
at the one-loop level result in effective
one-loop corrected values for the  Kobayashi-Maskawa (KM)
matrix elements~\cite{Babu,Isidori}.
These corrections modify the Wilson
coefficients involved in Eq.~(\ref{eq:braratio}), as can be seen 
in Eqs. (6.35) and (6.36) in ~\cite{Buras1} or in Eq. (14) in~\cite{Tata}.
We have included these flavour-violating effects as described 
in~\cite{Buras1,Tata}, taking into account the squark mixing effects. 

The counter-terms are mediated by $A,H,h$ exchange, as 
seen in Eqs. (5.1) and (5.2) of~\cite{Bobeth}, and dominate in the 
large-$\tan \beta$ limit, where the $\bmm$ decay amplitude $\propto 
1/m_{A}^2$ and hence the decay rate 
$\propto 1/m_{A}^4$.  This underlines the potential of $\bmm$ decay
for constraining models with large $\tan \beta$ and small $m_{A}$.

As already noted, in the NUHM $m_A$ may take values different from those 
required by the vacuum conditions in the CMSSM. Quite generally,
the electroweak vacuum conditions in the MSSM may be written in the form
\begin{equation}
\mu^2 \; = \; \frac{m_1^2 - m_2^2 \tan^2 \beta + \frac{1}{2}m_Z^2(1 - \tan^2 
\beta) 
+ \Delta_\mu^{(1)}}{\tan^2 \beta - 1 + \Delta_\mu^{(2)}} ,
\label{vac1}
\end{equation}
and
\begin{equation}
m_A^2(Q) \; = \; m_1^2(Q) + m_2^2(Q) + 2 \mu^2(Q) + \Delta_A(Q),
\label{vac2}
\end{equation}
and $m_{1,2} \equiv m_{1,2}(m_Z)$, where $\Delta_A, \Delta_\mu^{(1,2)}$ are 
loop
corrections~\cite{Pierce,Barger:1993gh,deBoer:1994he,Carena:2001fw,erz}.
The exact forms of the radiative corrections to $\mu$ and $m_A$ are not
needed for the discussion here, though we do note that the dominant contribution to
$\Delta_\mu^{(1)}$ at large $\tan \beta$ contains a term which is
proportional to $h_t^2 \tan^2 \beta$, whereas the dominant contribution to
$m_A^2$ contains terms proportional to $h_t^2 \tan \beta$ and $h_b^2 \tan
\beta$.  The radiative corrections between the values of the quantities
$\mu^2, m_{1,2}^2$ at $Q$ and the electroweak scale are well known. 

It is clear from (\ref{vac1}) and (\ref{vac2}) that departures of the
input supersymmetry-breaking contributions to $m_{1,2}^2$ from their
universal values in the CMSSM, as permitted in the NUHM, induce
corresponding changes in the allowed values of $\mu$ and $m_A^2$,
respectively. We evaluate all the relevant radiative corrections in our
analysis: the sensitivity of $m_A$ to $m_t$ and $m_b$ was discussed
extensively in~\cite{Bmumu}, and we do not discuss the issue any further
here.

Spin-independent elastic $\chi$-nucleon scattering is controlled by the
following effective four-fermion Lagrangian:
\begin{equation}
{\cal L} \, = \, \alpha_{3i} \bar{\chi} \chi \bar{q_{i}} q_{i},
\label{lagr}
\end{equation}
which is to be summed over the quark flavours $q$, and the
subscript $i$ labels up-type quarks ($i=1$) and down-type quarks
($i=2$). The model-dependent coefficients $\alpha_{3i}$ include terms
$\propto 1/m_{H_{1,2}}^2$~\cite{EFlO1}, where $H_{1,2}$ are the two scalar Higgs
bosons in the MSSM, in which it is well known that the lighter one, $H_2$, must have a mass
$\sim 120$~GeV, whereas the heavier one, $H_1$, has a mass very similar to
$M_A$. Hence the elastic cold dark matter scattering cross section also
increases for small $M_A$~\cite{DDnuhm}. The magnitude of the cross section depends on
hadronic matrix elements related to the $\pi$-nucleon $\Sigma$ term, for which
values between $\sim 64$ and $\sim 45$~MeV are frequently quoted. In the estimates
of the $\chi$-nucleon scattering used here, we assume $\Sigma = 64$~MeV,
which yields relatively large cross sections. Smaller regions of the NUHM
parameter space would be excluded if we used a smaller estimate of
$\Sigma$~\cite{DDupdate},
so this assumption maximizes the possible impact of the CDMS constraint.

\section{Analysis of NUHM Parameter Planes}

\begin{figure}
\vskip 0.5in
\vspace*{-0.75in}
%\hspace*{-.70in}
\begin{minipage}{8in}
\epsfig{file=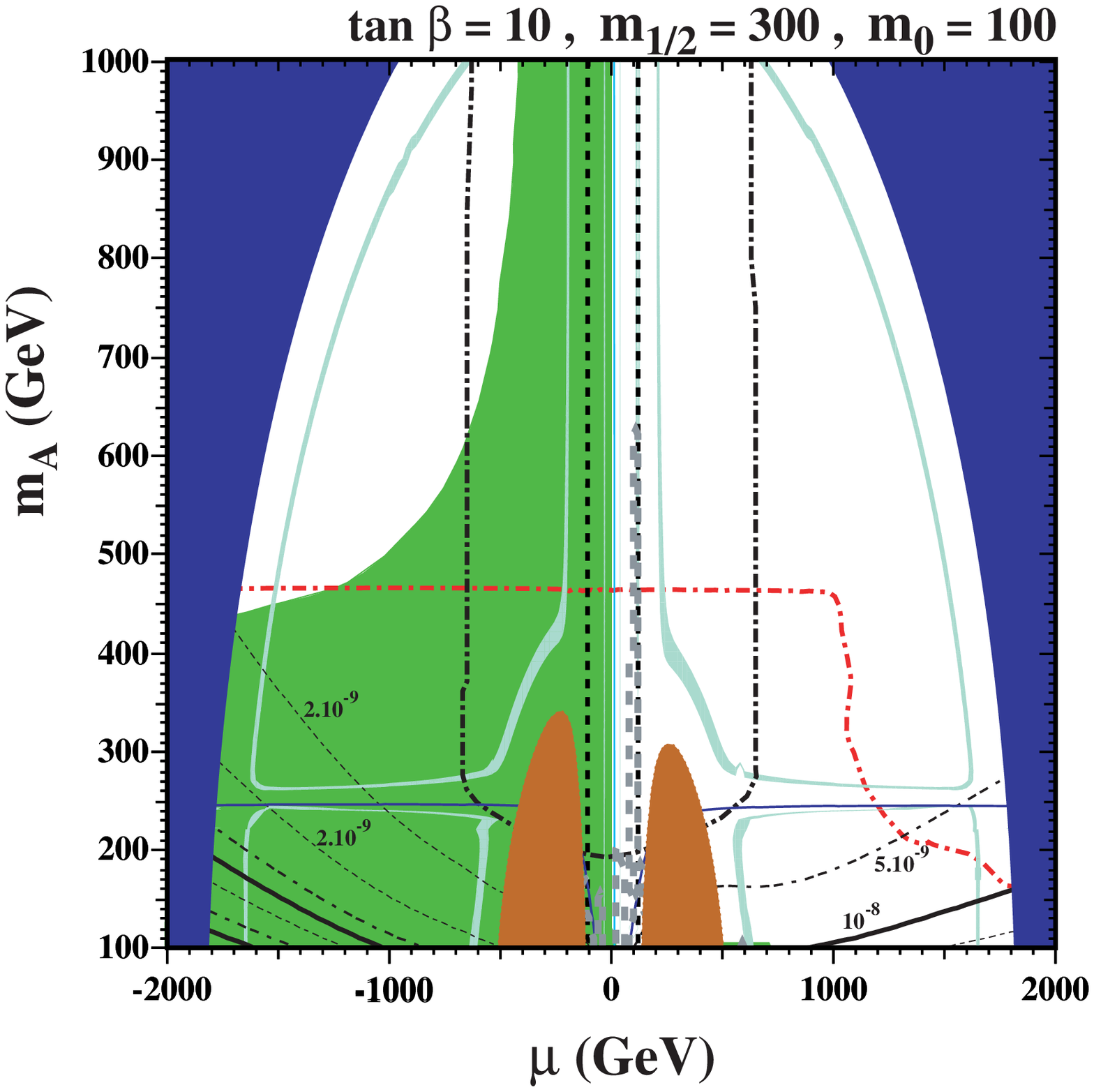,height=3.3in}
\hspace*{-0.17in}
\epsfig{file=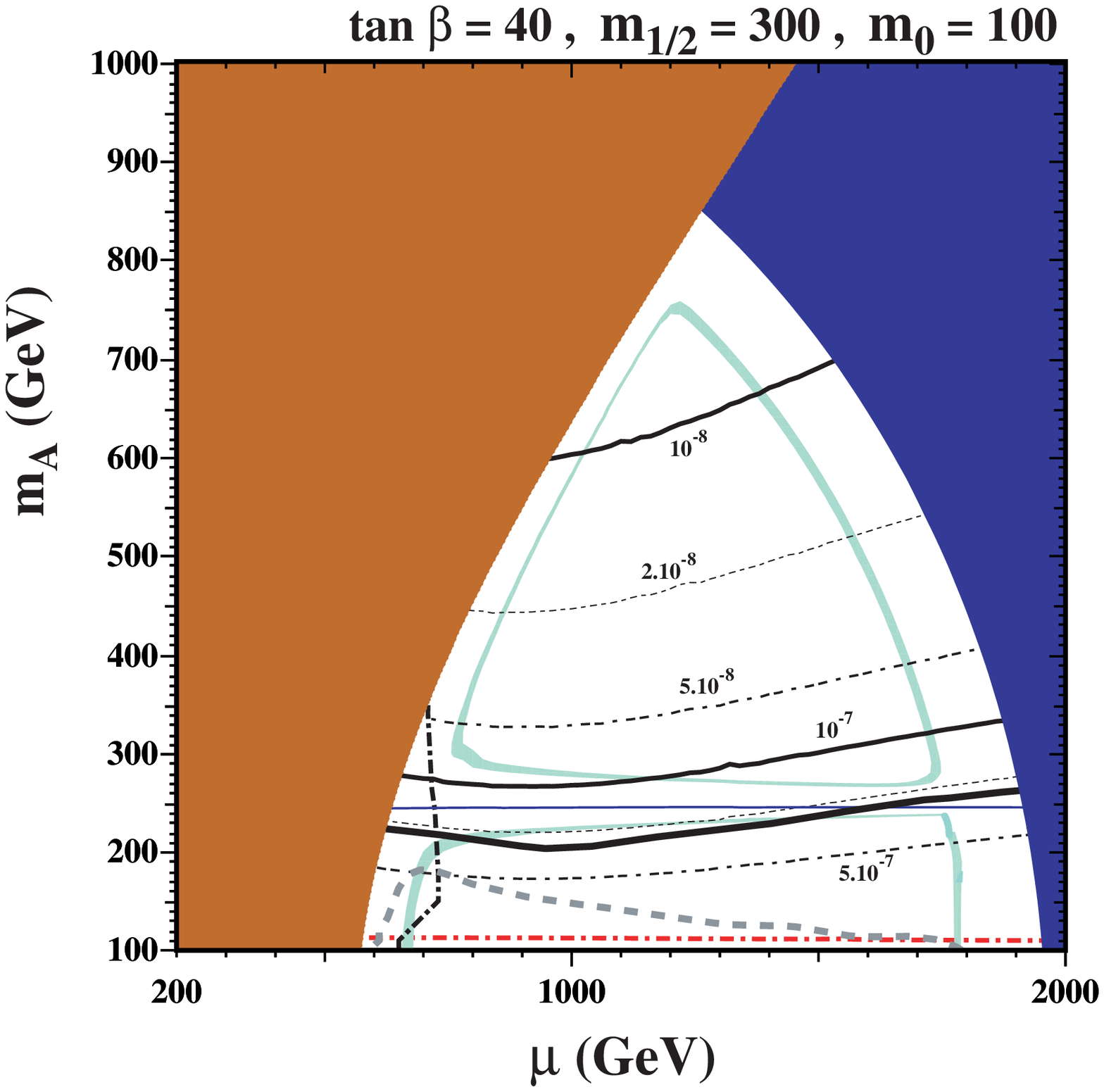,height=3.3in}
\hfill
\end{minipage}
\begin{minipage}{8in}
\epsfig{file=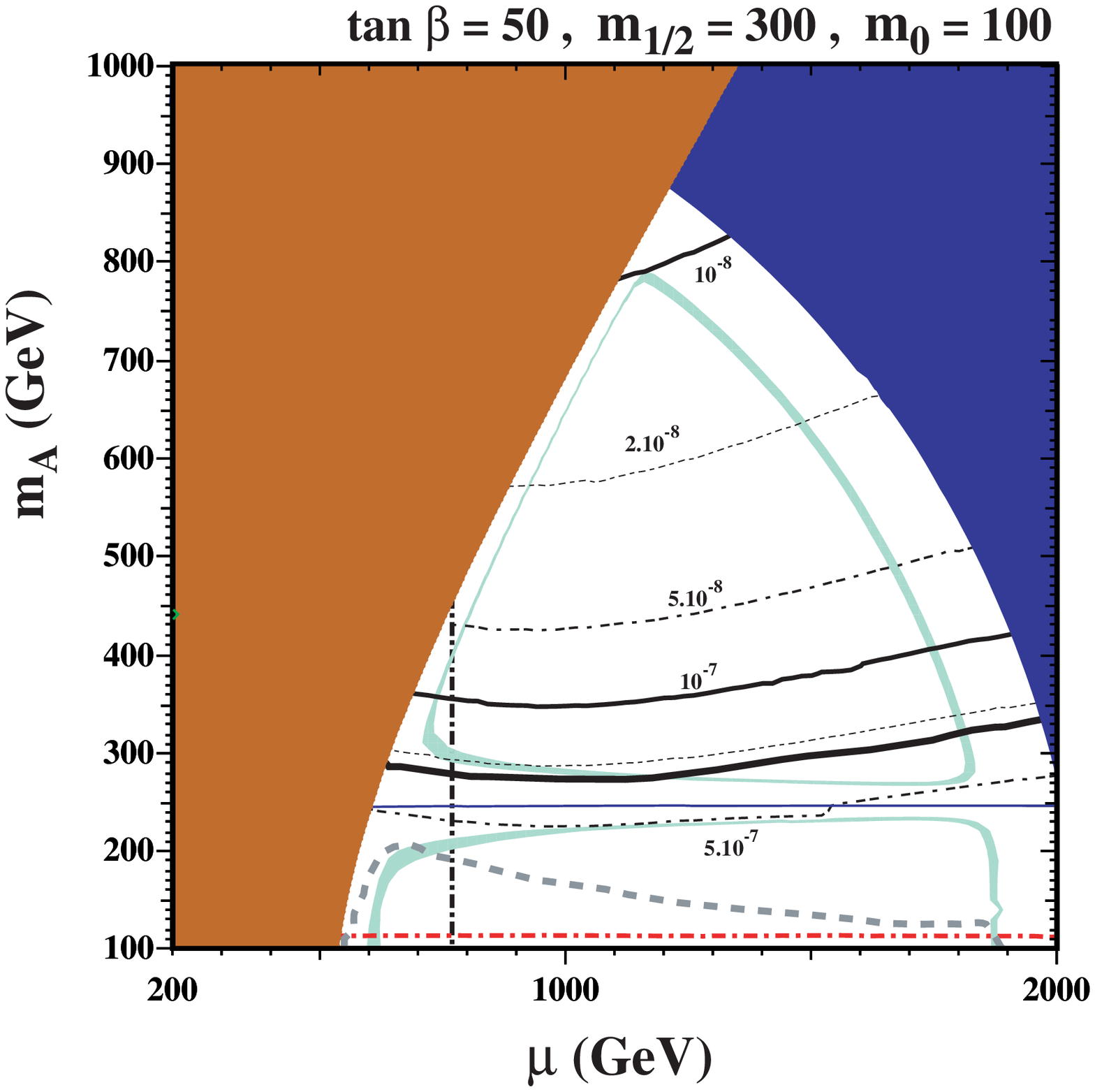,height=3.3in}
\hspace*{-0.17in}
\epsfig{file=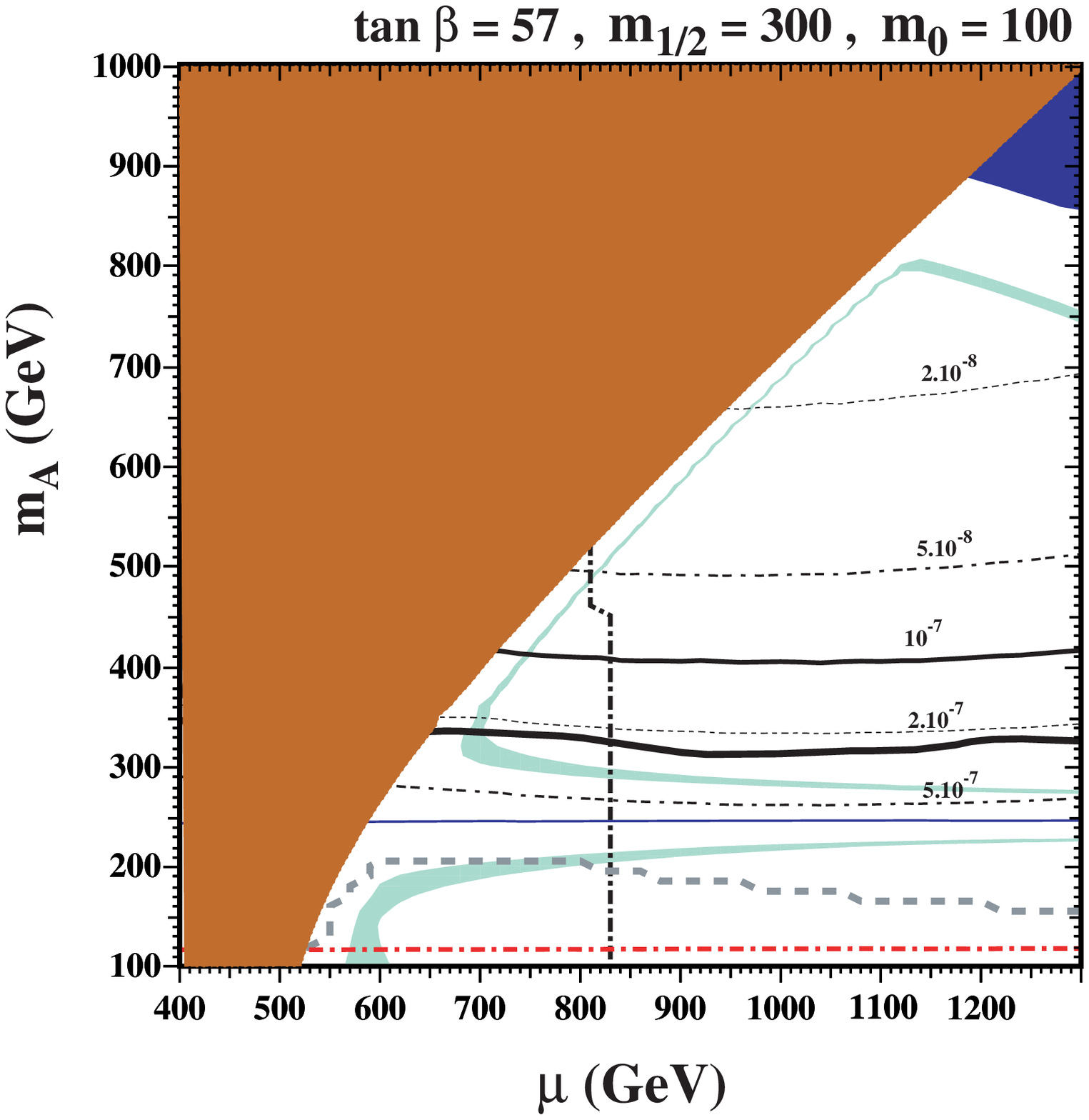,height=3.3in}
\hfill
\end{minipage}
\caption{
{\it
Allowed regions in the $(\mu, M_A)$ planes for $m_{1/2} = 300$~GeV and 
$m_0 = 100$~GeV, for (a) $\tan \beta = 10$, (b) $\tan \beta = 40$, (c) 
$\tan \beta = 50$ and (d) $\tan \beta = 57$. In each panel, the near-horizontal
solid blue line is the contour where $M_A = 2 m_\chi$, and the turquoise strips are 
those where the relic neutralino LSP density falls within the range 
favoured by WMAP and other cosmological and astrophysical observations. The 
LEP chargino limit is shown as a dashed black line and the GUT stability constraint as a dot-dashed black line. The regions disallowed by $b \to s \gamma$ are shaded green, and
those disallowed because the ${\tilde \tau}_1$ or the ${\tilde \nu_e}$
would be the LSP are shaded brick-red and dark blue, respectively.
The Higgs constraint is the largely horizonal red dot-dashed line. 
Contours of the $\bmm$ branching ratio are labelled correspondingly, with the
current Tevatron limit the boldest black line, and the
CDMS constraint is shown as a thick dashed grey line.
In panel a), the $g_\mu-2$ constraint is satisfied when $\mu > 0$. In the remaining 
three panels,  the contribution to $g_\mu-2$ is too large at the 2-$\sigma$ level.
}}
\label{fig:300_100} 
\end{figure}

In order to exemplify the possible effects of the $\bmm$ and CDMS
constraints, we display some specific NUHM $(\mu, m_A)$ planes for
different values of $\tan \beta$, $m_{1/2}$ and $m_0$, exhibiting the
interplay of the different experimental, phenomenological and theoretical
constraints. We first consider the case $\tan \beta = 10$, $m_{1/2} =
300~{\rm GeV}$ and $m_0 = 100$~GeV, shown in panel (a) of
Fig.~\ref{fig:300_100}. This value of $\tan \beta$ is towards the lower
end of the range where we find generic CMSSM models satisfying all the
constraints, for both signs of $\mu$, except for the $g_\mu-2$ constraint that
is satisfied only for positive $\mu$. 
The most important constraints for $\mu > 0$ are those due to the chargino mass, shown as a vertical black
dashed line at low $\mu \sim 100$~GeV, and the GUT stability
constraint at larger $\mu \sim 650$~GeV, shown as a near-vertical black
dash-dotted line that turns horizontal at low $m_A \sim 250$~GeV. Regions between these lines
are consistent with both constraints, but we also note a small excluded (brick-red) `sugarloaf' around $\mu \sim
300$~GeV that extends up to $M_A \sim 300$~GeV, where the LSP would have
been the lighter stau, ${\tilde \tau}_1$. For $\mu < 0$, a large region is excluded
by $b \to s \gamma$ decay, as shown by the green (medium) shading here 
and in subsequent figures. The excluded regions where the ${\tilde \nu_e}$ 
would be the LSP (or become tachyonic) are shaded (dark) blue: in this panel, they lie beyond
the GUT stability region. The red dash-dotted line marks the LEP Higgs
constraint~\footnote{This is evaluated following the likelihood approach
described in~\cite{EHOW4}.}, which in this case excludes the lower part of the region
allowed by GUT stability, and requires $M_A \ga 460$~GeV. 
Finally, in this panel, 
regions with $\mu > 0$
are allowed by $g_\mu -2$~\footnote{We assume $\delta a_\mu$, where $a_\mu \equiv
(g_\mu-2)/2$, to be within the range 6.8 to
43.6 $\times 10^{-10}$ at the 2-$\sigma$ level.}, whereas regions with $\mu < 0$ are
disallowed at the 2-$\sigma$ level.

Within the region allowed by the other constraints, we note that there is
a near-vertical WMAP strip extending upwards. All of this WMAP strip is allowed
comfortably by the current $\bmm$ constraint, since the branching ratio
for $\bmm$ is very close to its Standard Model value $\sim 3.9 \times
10^{-9}$. It is dubious whether even the LHC would be able to distinguish
the NUHM from the Standard Model for these values of $\tan \beta$, $m_{1/2}$
and $m_0$. 
Other regions with acceptable relic densities either have a small Higgs mass
or violate the GUT constraint.  These include the lower strip which extends upwards from
$M_A = 100$~GeV and 
turns towards the horizontal at lower $M_A$, where rapid $\chi \chi \to A,
H$ annihilation becomes important when $M_A \sim 2 m_\chi$
(near-horizontal solid blue line) and the WMAP strip in the upper left which is
determined by neutralino-sneutrino co-annihilations.
Also shown as a dashed grey line is the constraint imposed by
the CDMS upper limit on spin-independent elastic cold dark matter
scattering. Here and in the subsequent figures, in regions of the NUHM
parameter space where the calculated relic LSP density $\Omega_\chi$ falls
below the WMAP range, the cross section is rescaled by the factor
$\Omega_\chi / \Omega_{CDM}$, in order to compensate for the the fact
that neutralinos could provide only this fraction of the galactic halo. As
a result, in this case, not only does the CDMS limit not exclude any of
the WMAP strip, it also does not exclude models surviving the LEP chargino
constraint.

When $\tan \beta = 40$~\footnote{For this and higher values of $\tan
\beta$, we find consistent electroweak vacua mostly only for positive values of
$\mu$.}, the stau constraint plays a much more important role, as seen in
panel (b). It excludes all the parameter space above and to the left of a
diagonal line that meets the GUT stability constraint, which is
essentially unchanged, at $\mu \sim 700$~GeV and $M_A \sim 350$~GeV. The
${\tilde \nu_e}$ constraint appears at much larger $\mu$ and $M_A$. In
this and subsequent panels of this figure, the LEP Higgs constraint
excludes only a narrow strip below $M_A \sim 120$~GeV, whereas $(g - 2)_\mu$ is 
incompatible with the measurement at the 2-$\sigma$ level. The only region
allowed by the GUT stability constraint and by WMAP is then a short strip
with $\mu \sim 700$~GeV, which lies above the Higgs constraint and
below the solid blue line where $m_A = 2 m_\chi$. {\it All of this strip
is excluded by the current $\bmm$ constraint} (represented by bold solid line). 
We see also that the CDMS
constraint excludes a region of the WMAP strip that is, however, already
excluded by the stronger $\bmm$ constraint in this case. As already
mentioned, the CDMS limit is rescaled when the calculated $\Omega_\chi$
falls below the WMAP range, which is responsible for the droop in the
dashed grey line for $\mu \sim 600$~GeV. In this and subsequent figures,
the CDMS limit is also shown in regions where $\Omega_\chi$ exceeds the
WMAP range, even though this region is disallowed by cosmology.

In panel (c) for $\tan \beta =
50$, the stau LSP constraint meets the GUT stability constraint at a somewhat higher values of
$M_A \sim 450$~GeV. The region by WMAP and GUT stability is now bisected by the $M_A = 2
m_\chi$ line, with strips both above and below this line. {\it The lower strip,
where $m_\chi > M_A/2$, is already excluded by the present $\bmm$
constraint}, and the upper strip lies only just beyond the present
sensitivity. The lower WMAP strip would also be almost excluded by the CDMS
constraint. Finally, in panel (d) for $\tan \beta = 57$, which is close
to the maximum value for which we find generic solutions to the
electroweak vacuum conditions, we see that the stau LSP constraint now intersects the GUT stability constraint at $M_A \sim 550$~GeV, with the ${\tilde \nu_e}$ LSP constraint appearing only at larger values of $\mu$. There
are again two strips allowed by WMAP, above and below the line where $M_A
= 2 m_\chi$. However, {\it in this case, as well as the lower WMAP strip, a large portion of the upper strip is also already excluded by
$\bmm$}, reflecting the greater power of this constraint as $\tan \beta$
increases. In this case, the CDMS constraint almost excludes the lower WMAP strip.

\begin{figure}
\vskip 0.5in
\vspace*{-0.75in}
%\hspace*{-.70in}
\begin{minipage}{8in}
\epsfig{file=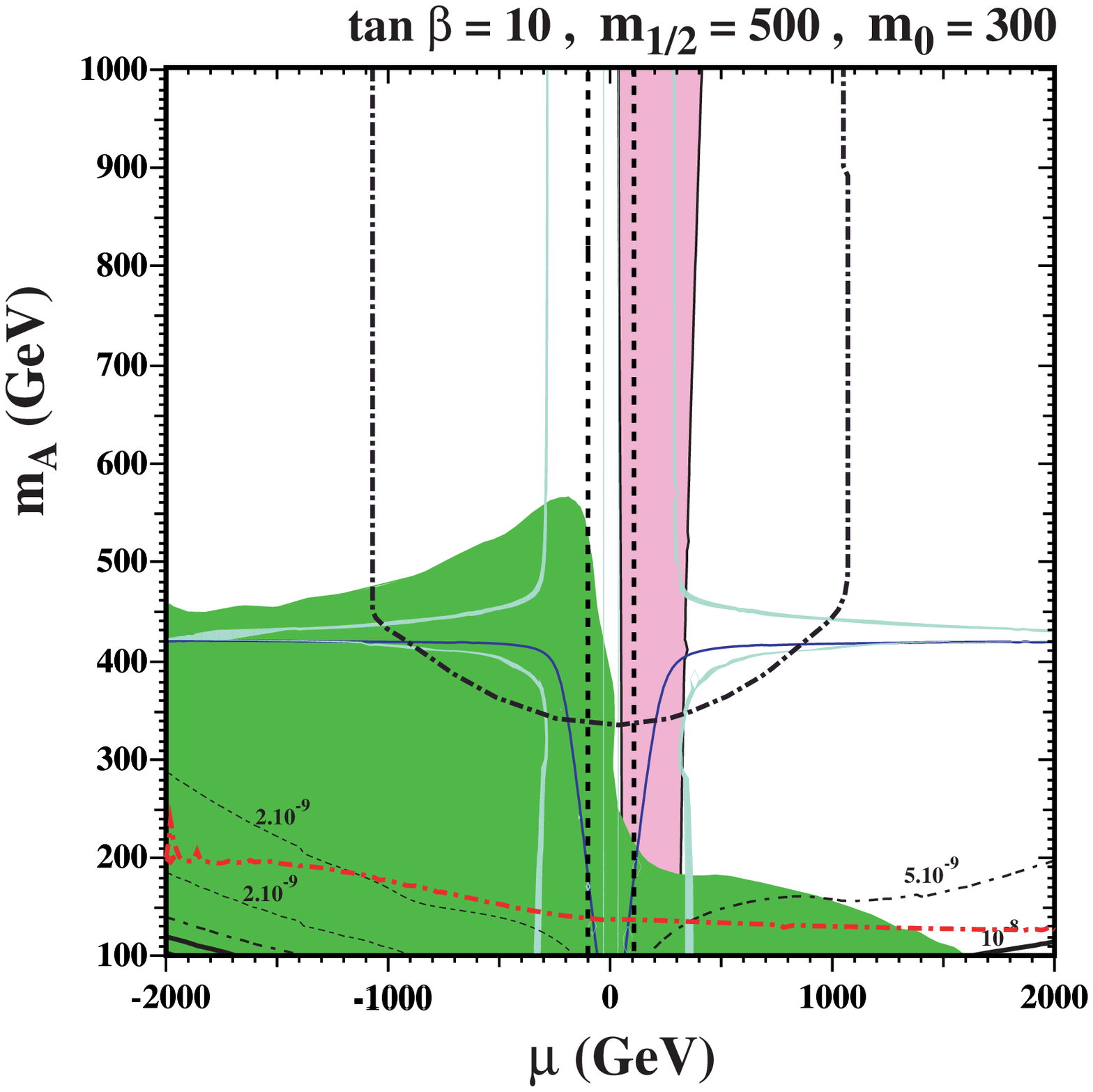,height=3.3in}
\hspace*{-0.17in}
\epsfig{file=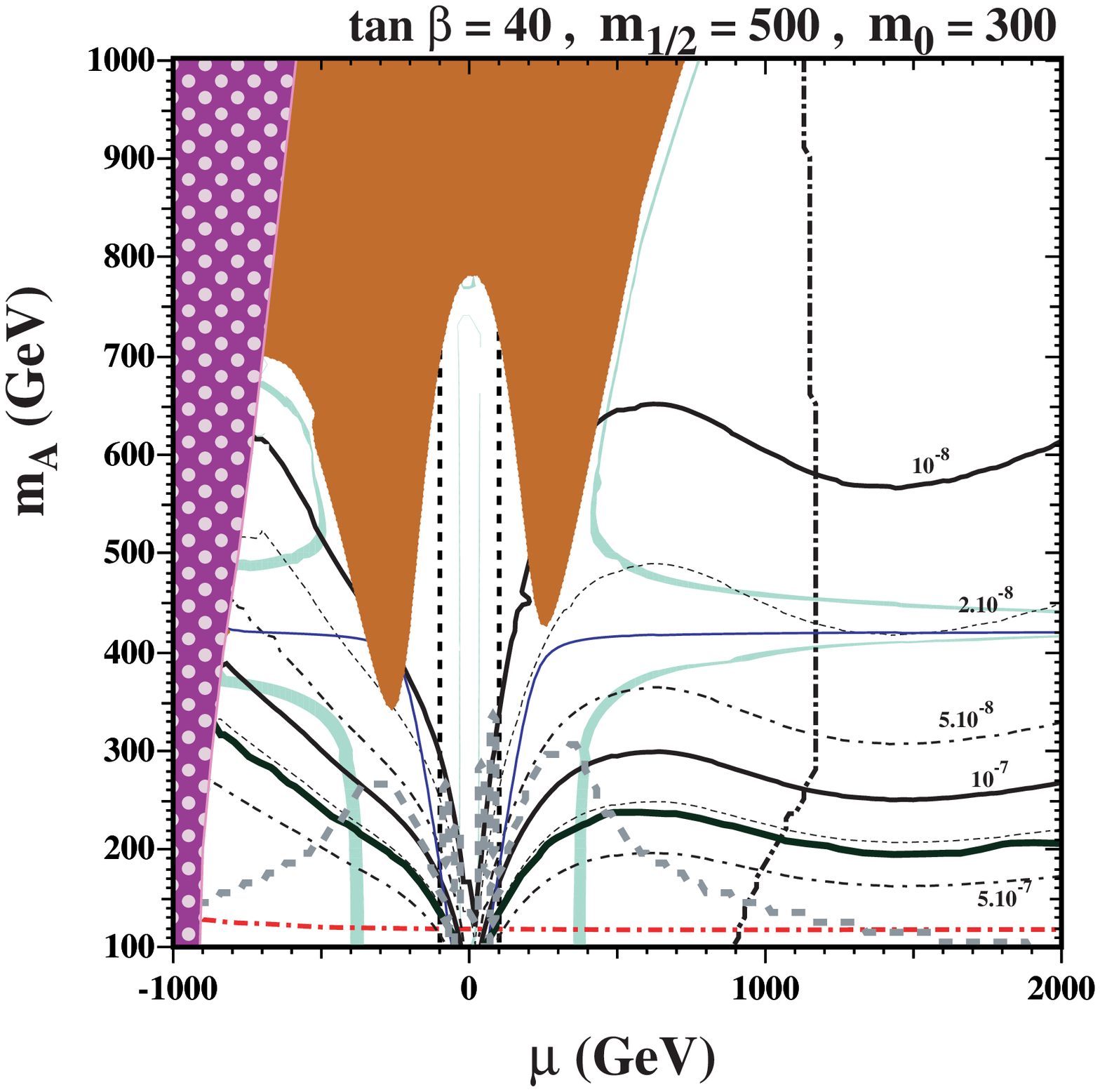,height=3.3in}
\hfill
\end{minipage}
\begin{minipage}{8in}
\epsfig{file=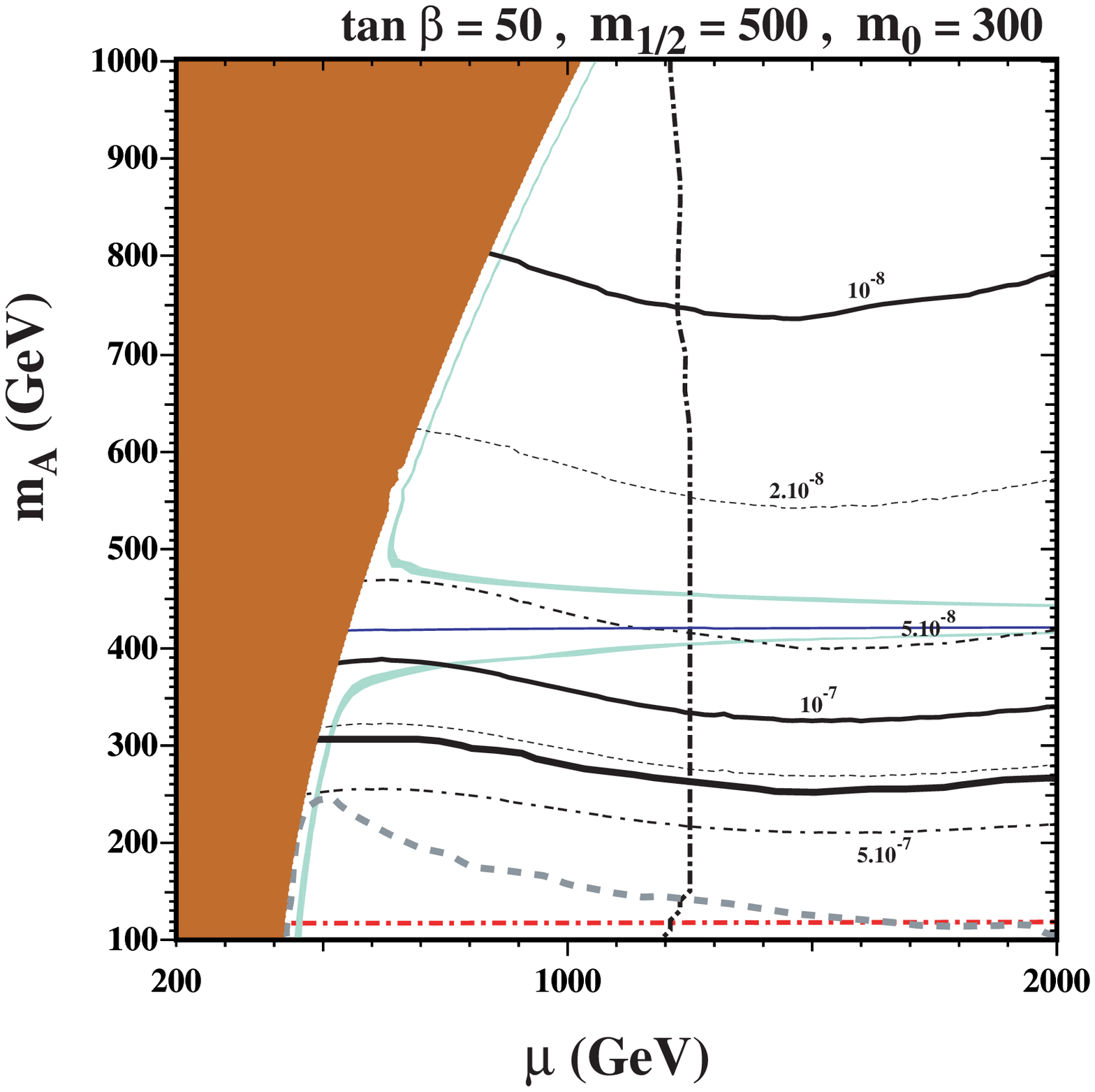,height=3.3in}
\hspace*{-0.17in}
\epsfig{file=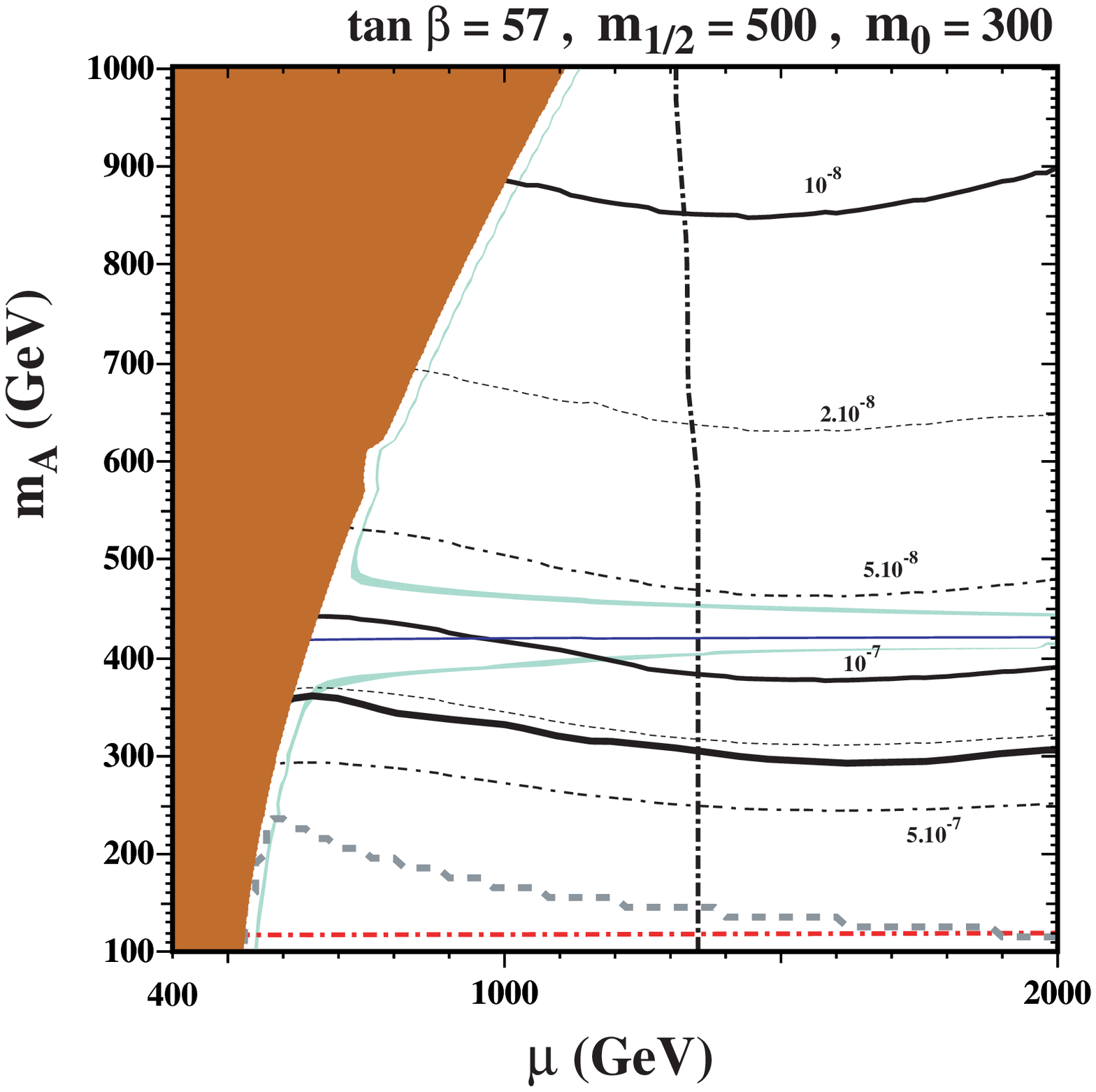,height=3.3in}
\hfill
\end{minipage}
\caption{
{\it
As in Fig.~\protect\ref{fig:300_100}, for the same values of $\tan 
\beta$ but for the choices $m_{1/2} = 500$~GeV and $m_0 = 300$~GeV. 
There is no electroweak symmetry breaking in the polka-dotted region of panel (b).
In panel a), the vertical pink shaded region shows the region of $g_\mu - 2$ allowed at the 2-$\sigma$ level.  In the remaining three panels, $g_\mu -2$ is allowed when $\mu > 0$. }}
\label{fig:500_300}
\end{figure}

Some qualitatively similar features are seen in Fig.~\ref{fig:500_300},
which displays analogous panels for the case $m_{1/2} = 500$~GeV, $m_0 =
300$~GeV. For $\tan \beta = 10$, both signs of $\mu$ are equally 
possible, whereas we find no consistent electroweak vacuum in the 
polka-dotted region for $\tan \beta = 40$, and no solutions with $\mu < 0$ for the larger values of $\tan \beta$. In all the panels, the GUT stability constraint provides the
right boundary of the allowed region at $\mu \sim 1100$ to 1350~GeV and, in
panel (a) also a lower limit on $M_A \sim 350$~GeV. For $\mu > 0$, the
LEP constraint on the chargino mass again supplies the left boundary in
panel (a) for $\tan \beta = 10$~\footnote{In panel (a), regions outside the
near-vertical (pink) band at $\mu \sim 200$~GeV are excluded by $g_\mu -2$. In the remaining three panels, 
all regions shown with $\mu > 0$ are  allowed by $g_\mu - 2$. In both panels (a)
and (b) there are regions where $\mu < 0$ that are allowed by the other
constraints, but disallowed by $g_\mu -2$.}, and partially in panel (b) for $\tan
\beta = 40$. The rest of the left boundary for $\mu > 0$ in panel (b), and the entire
left boundaries in panels (c, d) for $\tan \beta = 50, 57$, respectively,
are provided by the stau LSP constraint. The bottom boundaries of the
allowed regions in panels (b, c, d) are provided by the LEP Higgs
constraint.

As in panels (c, d) of Fig.~\ref{fig:300_100}, each panel features a pair
of WMAP strips, above and below the $M_A = 2 m_\chi$ line.  When $\tan
\beta = 10$, the entire WMAP strips are allowed by $\bmm$, and sensitivity
to the Standard Model prediction would be required to challenge any part
of them. However, {\it in the remaining panels, increasing parts of the lower
WMAP strips are excluded by $\bmm$ as $\tan \beta$ increases}.  However, in
each case sensitivity to $\bmm$ below $10^{-8}$ would be required to
explore all of the upper WMAP strip. The CDMS constraint lies outside the
GUT stability region for $\tan \beta = 10$. On the other hand, it excludes
a somewhat larger part of the lower WMAP strip than does $\bmm$ for $\tan
\beta = 40$, whereas the $\bmm$ constraint is stronger for $\tan \beta =
50$ and 57.

\begin{figure}
\vskip 0.5in
\vspace*{-0.75in}
%\hspace*{-.70in}
\begin{minipage}{8in}
\epsfig{file=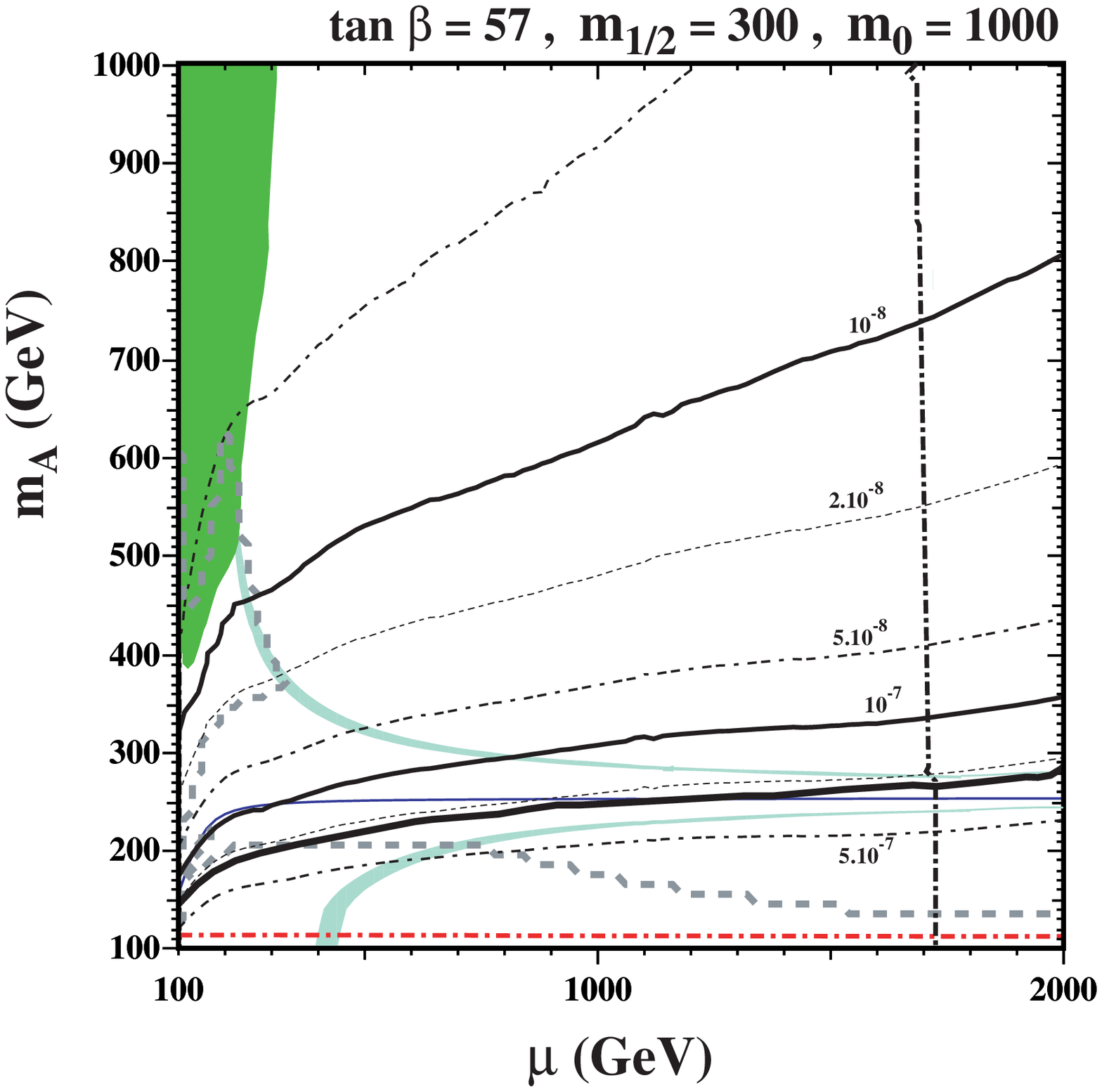,height=3.3in}
\hspace*{-0.17in}
\epsfig{file=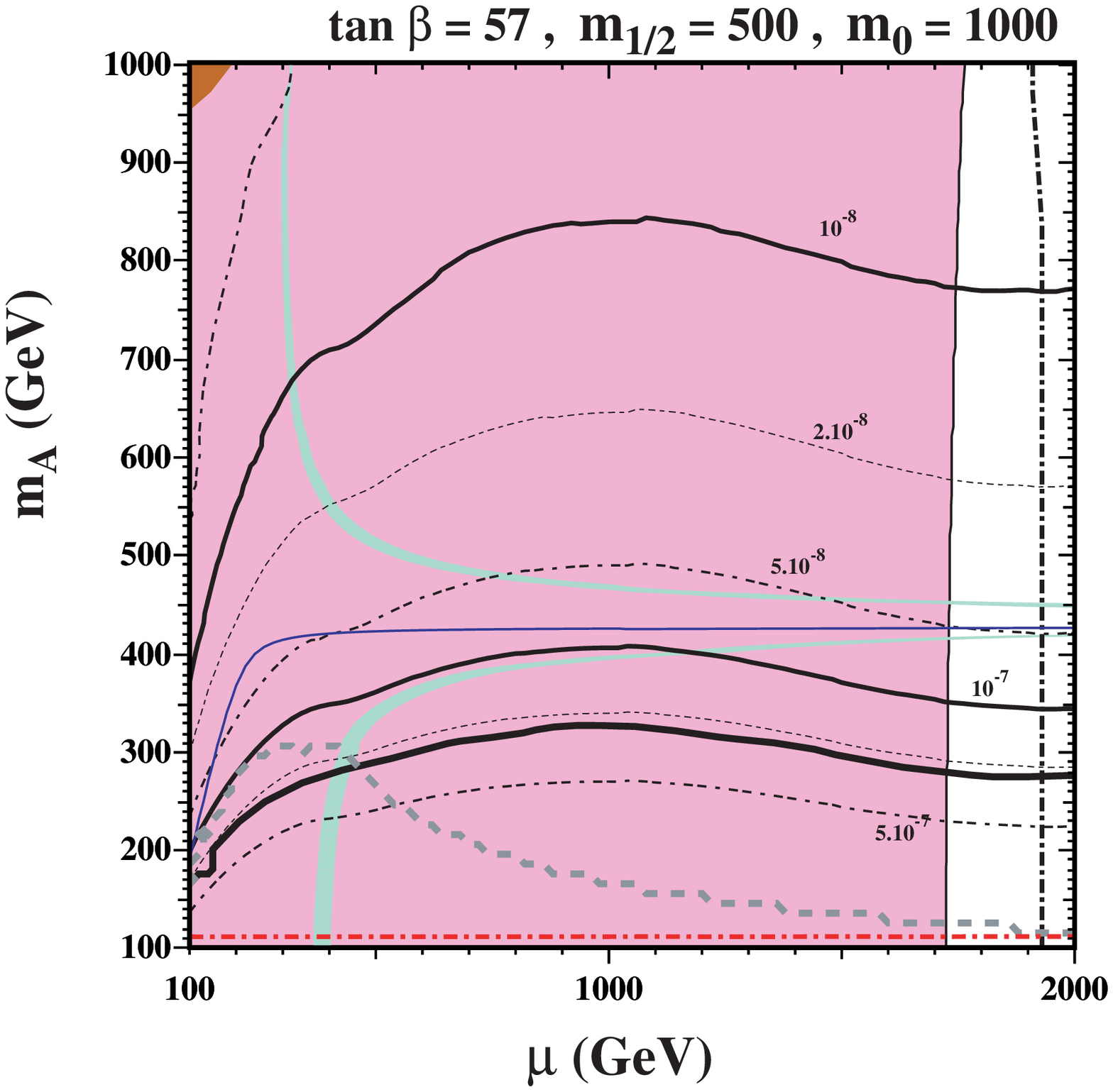,height=3.3in}
\hfill
\end{minipage}
\begin{minipage}{8in}
\epsfig{file=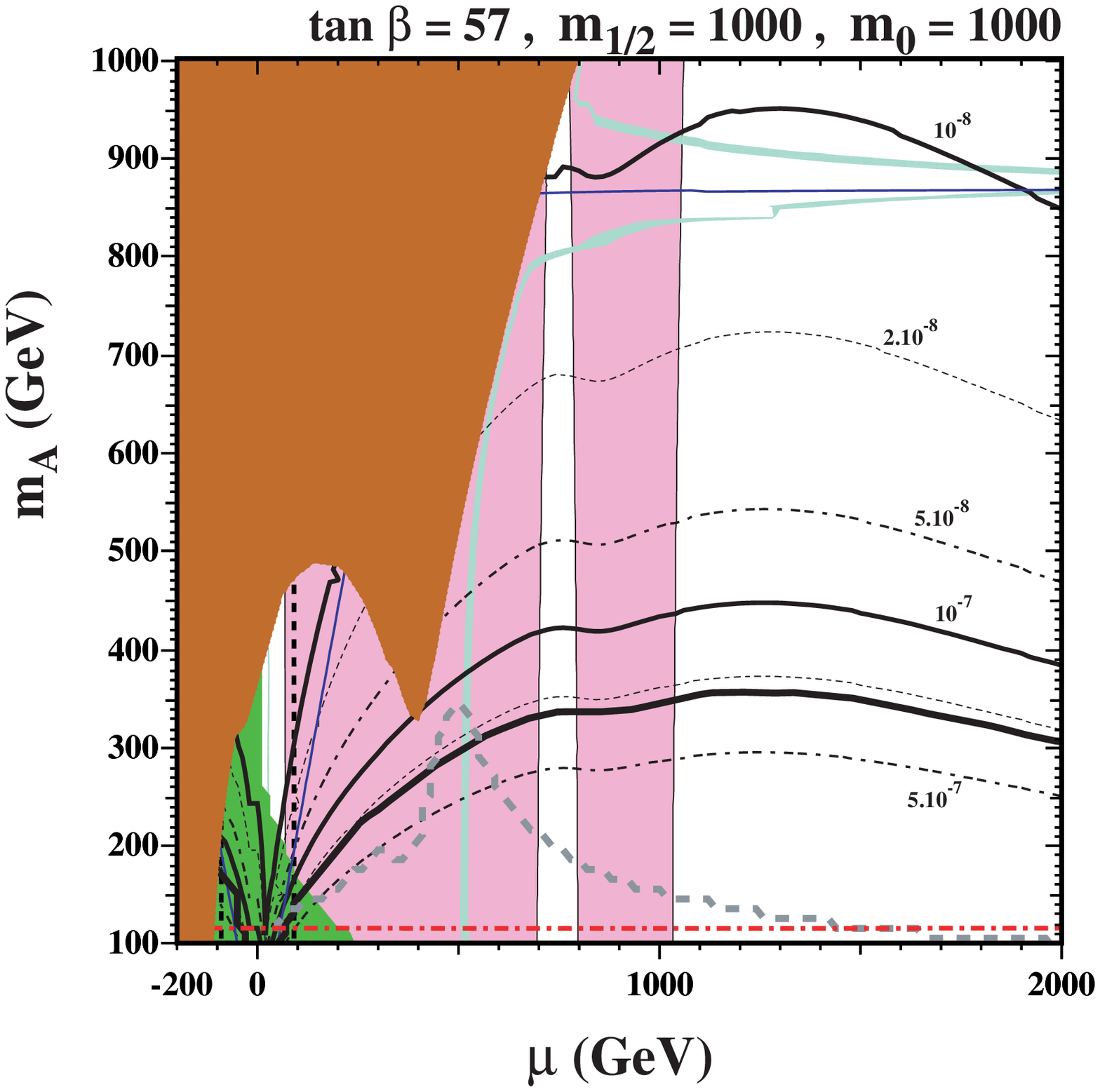,height=3.3in}
\hspace*{-0.17in}
\epsfig{file=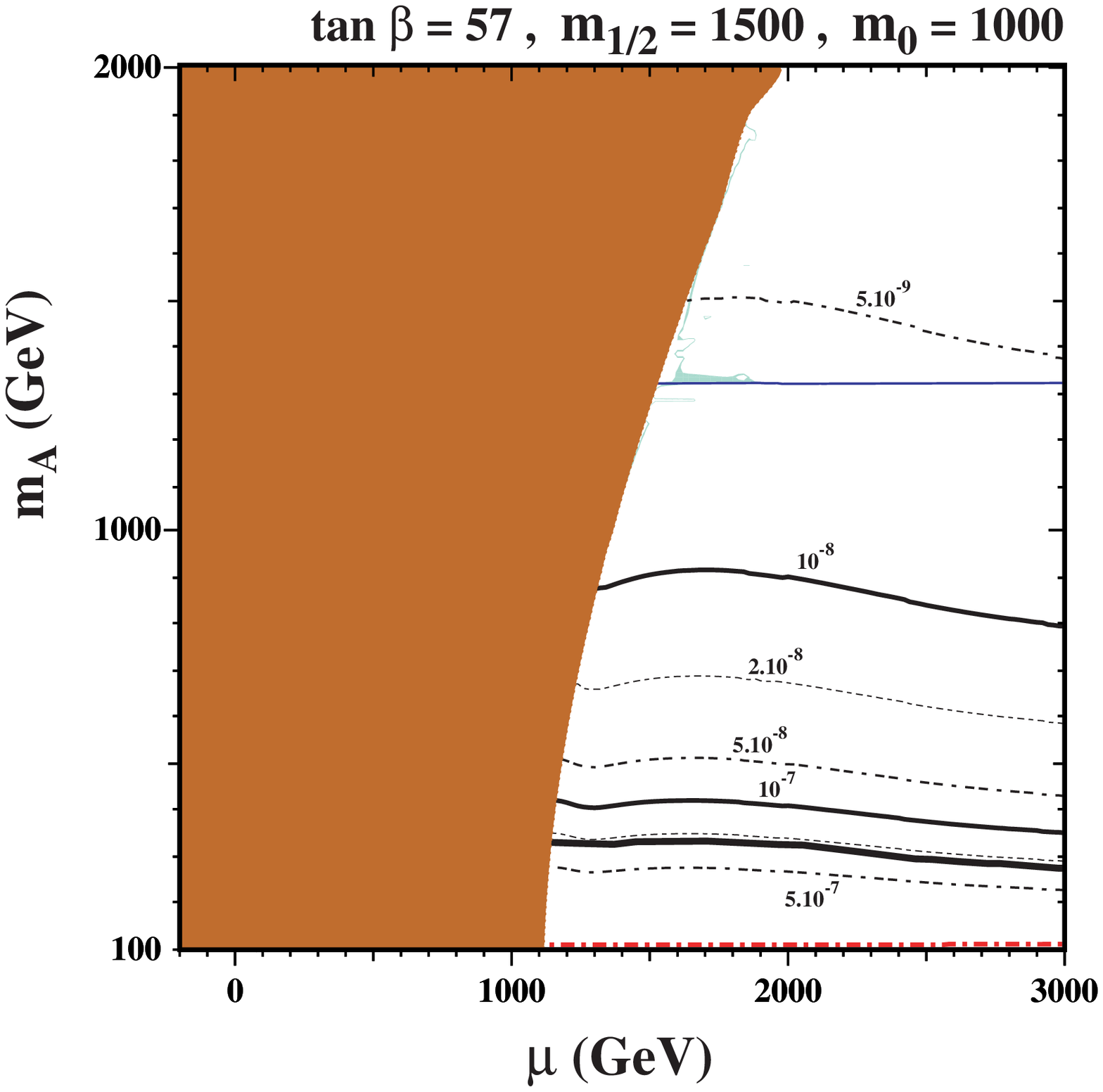,height=3.3in}
\hfill
\end{minipage}
\caption{
{\it
As in Fig.~\protect\ref{fig:300_100}, but for the choices $m_{1/2} =$ (a) 
300~GeV, (b) 500~GeV, (c) 1000~GeV and (d) 1500~GeV, all for $\tan 
\beta = 57$ and $m_0 =  1000$~GeV. The 
regions allowed by the $g_\mu - 2$ constraint are shaded pink (light grey) in 
panels (b) and (c). In panel (a), the region with $\mu > 0$ is allowed by $g_\mu
-2$, whereas that in panel (d) is disallowed at the 2-$\sigma$ level.}}
\label{fig:57}
\end{figure}

Both the stau LSP and GUT stability constraints weakened between
Figs.~\ref{fig:300_100} and \ref{fig:500_300}, the culprit being the
increased value of $m_0$, which makes the stau heavier, and also makes the
vacuum more likely to be stable. Therefore, in order to display more
clearly the interplay of $\bmm$ with the line where $M_A = 2 m_\chi$ and
the pairs of upper and lower WMAP strips, we now consider a series of
cases with relatively large $m_0 = 1000$~GeV, for which the stau and GUT
stability constraints are irrelevant for large ranges of $\mu$ and $M_A$.  
Panels (a, b, c, d) of Fig.~\ref{fig:57} show the cases $m_{1/2} = 300,
500, 1000$ and 1500~GeV, respectively, all for $\tan \beta = 57$. The regions
allowed by $(g - 2)_\mu$ at the 2-$\sigma$ level are shown explicitly in panels (b) and (c): at this level, the entire $\mu > 0$ region is allowed in panel (a) and disallowed in panel (d). We
notice that the lines where $M_A = 2 m_\chi$ move upwards as $m_{1/2}$
increases, reflecting the fact that $m_\chi \propto m_{1/2}$,
approximately. On the other hand, the line representing the current upper
bound on $\bmm$ is relatively insensitive to both $m_{1/2}$ and $\mu$. For
this reason, whereas {\it $\bmm$ already excludes the lower WMAP strip for the
choice $m_{1/2} = 300$~GeV} in panel (a) [compare also its impacts in
panels (d) of Figs.~\ref{fig:300_100} and \ref{fig:500_300}], {\it $\bmm$ is
currently sensitive to only progressively smaller fractions of the lower
WMAP strip as $m_{1/2}$ increases} in panels (b, c, d). By comparison, the CDMS constraint
excludes only part of the lower WMAP strip in panel (a), but also a portion of the upper 
WMAP strip at small $\mu$.
The peculiar shape of the CDMS curve in panel a) is caused by our scaling
to the relic density, which is particularly important when $m_\chi \approx M_A/2$.
 In panels (b) and (c), CDMS excludes slightly more of the lower WMAP strip than does $\bmm$, but this advantage would be removed if one adopted the lower value $\Sigma = 45$~MeV. The CDMS constraint has no significant impact in panel (d).

To summarize our findings on the current impact of $\bmm$ in the $(\mu, M_A)$ 
planes: its importance increases with $\tan \beta$, and for large values it 
may exclude substantial parts of the WMAP strip where $M_A < 2 m_\chi$. 
The impact of $\bmm$ does not vary rapidly with $m_0$, but is relatively 
less important as $m_{1/2}$ increases.

\begin{figure}
\vskip 0.5in
\vspace*{-0.75in}
%\hspace*{-.70in}
\begin{minipage}{8in}
\epsfig{file=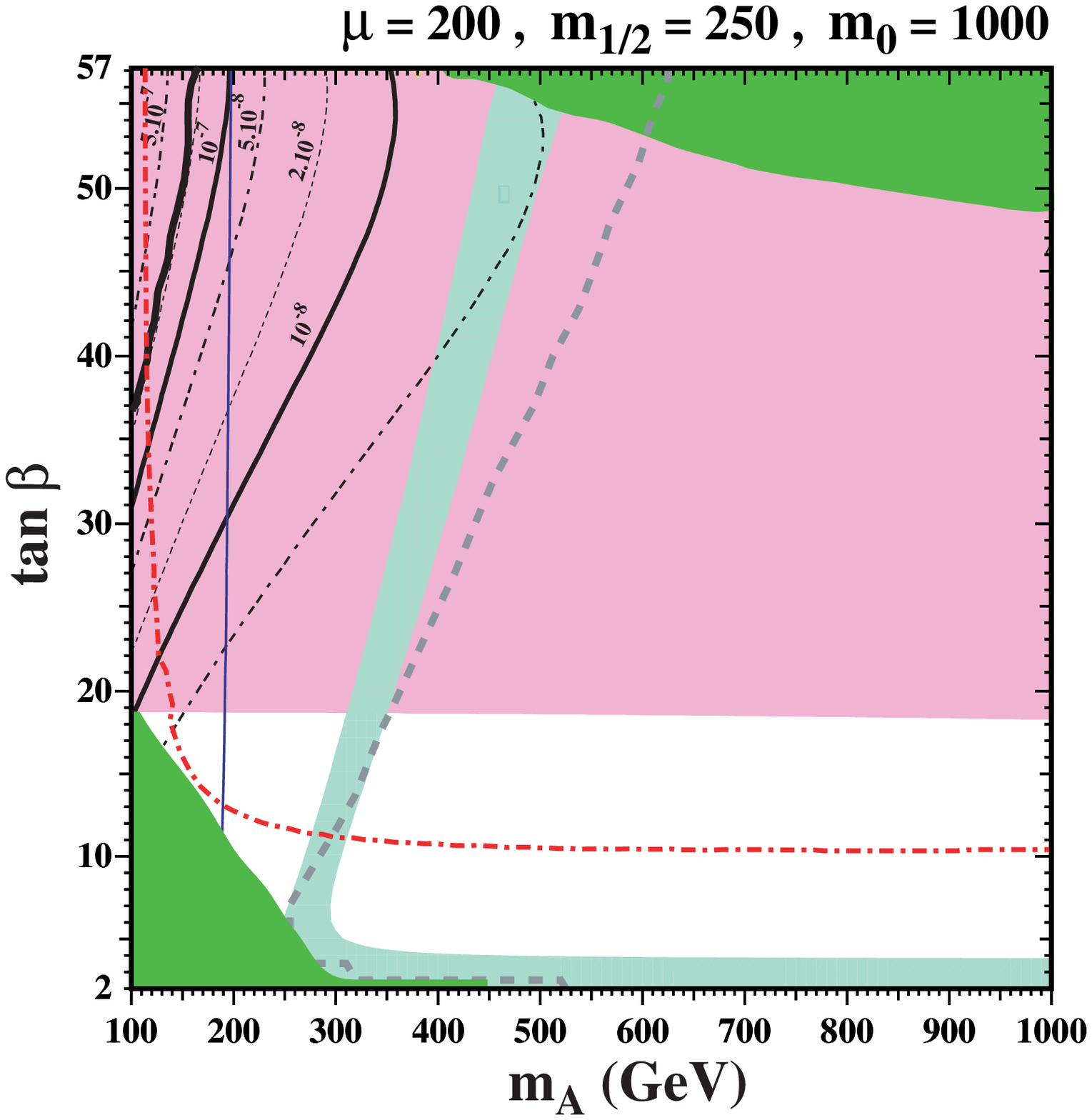,height=3.3in}
\hspace*{-0.17in}
\epsfig{file=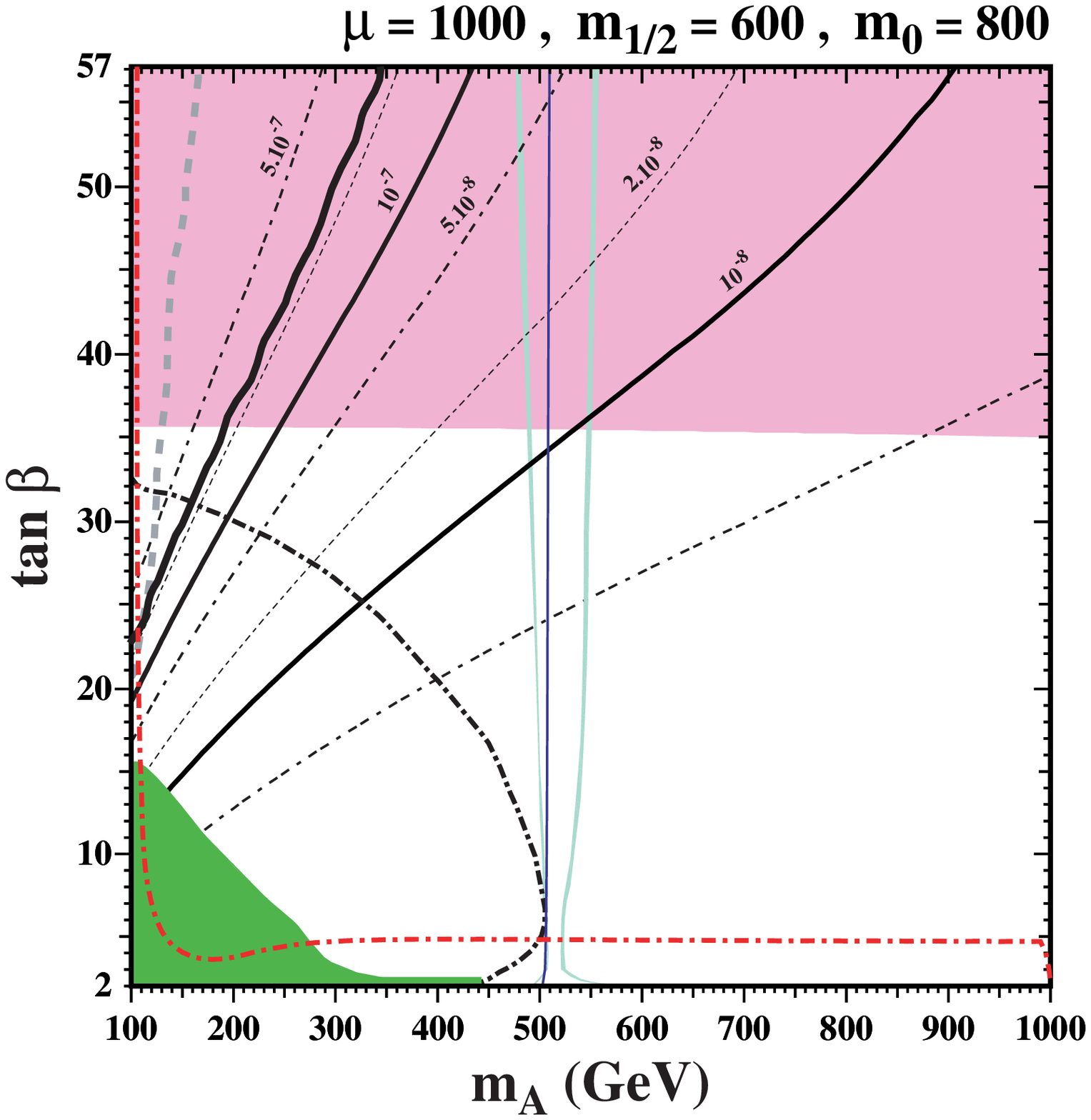,height=3.3in}
\hfill
\end{minipage}
\caption{
{\it
Allowed regions in the $(M_A, \tan \beta)$ planes for (a) $\mu = 1000$~GeV, $m_{1/2} = 
600$~GeV and $m_0 = 800$~GeV and (b) $\mu = 800$~GeV, $m_{1/2} = 
250$~GeV and $m_0 = 1000$~GeV. The constraints are displayed in the same way as in
in Fig.~\protect\ref{fig:300_100}.
}}
\label{fig:mAtb}
\end{figure}

We now turn to some examples of parameter $(M_A, \tan \beta)$ planes,
shown in Fig.~\ref{fig:mAtb}, which represent orthogonal projections of the NUHM parameter 
space that are often favoured in analyses of MSSM Higgs phenomenology~\cite{MAtb}. For
convenience of comparison, our examples are taken from~\cite{nuhm}. We see in
panel (a) for $\mu = 200$~GeV, $m_{1/2} = 250$~GeV and $m_0 = 1000$~GeV
that $b \to s \gamma$ excludes regions at both large and small values of 
$(M_A, \tan \beta)$. The strip allowed by WMAP is relatively broad, with a portion
to the right of the blue line where $M_A = 2 m_\chi$, and another part at low 
$\tan \beta$, but the latter is excluded by the LEP Higgs constraint. Only
the parts of the plane with $\tan \beta \ga 19$ are allowed by $g_\mu -2$. 
{\it In this
case, only a small region at small $M_A$ and large $\tan \beta$ is excluded by
$\bmm$. On the other hand, the CDMS experiment excludes all parts of the WMAP 
strip that are allowed by $g_\mu -2$}~\footnote{This latter conclusion would, however, no longer hold if one adopted $\Sigma = 45$~MeV.}.

Turning now to panel (b) of Fig.~\ref{fig:mAtb} for $\mu = 1000$~GeV, $m_{1/2} = 
600$~GeV and $m_0 = 800$~GeV, we see that $b \to s \gamma$ now excludes
only a region at small values of $(M_A, \tan \beta)$, and that the requirement of 
vacuum stability up to the GUT scale also excludes a region at $M_A \la 500$~GeV. 
The WMAP strip is now much narrower than in panel (a), with portions on either 
side of the line where $M_A = 2 m_\chi$. The LEP Higgs constraint is weaker, excluding only
$\tan \beta \la 5$, whereas the $g_\mu -2$ constraint is stronger, excluding the region
with $\tan \beta \la 35$. In this case, neither the $\bmm$ nor the dark matter
scattering constraints have any impact on the allowed portions of the WMAP strips at
larger $\tan \beta$.

As already mentioned, $(M_A, \tan \beta)$ planes are often considered in discussions of
MSSM Higgs phenomenology. These two examples show that only small parts of
these planes may be allowed by the various theoretical, phenomenological and
cosmological constraints. In the particular examples studied, $\bmm$ is not yet an
important constraint, whereas the search for astrophysical dark matter may be.
Generic $(M_A, \tan \beta)$ regions may
be allowed if one exploits the flexibility of the NUHM to vary $\mu$, $m_{1/2}$
and $m_0$
independently at fixed $M_A$ and $\tan \beta$. We plan to return to a more detailed 
discussion of $(M_A, \tan \beta)$ planes in the future.

\section{Summary and Prospects}

We have shown in this paper that the current  $\bmm$ constraint already imposes
significant constraints on the NUHM, excluding generic regions with small $M_A$
and large $\tan \beta$ that would have been allowed by the other theoretical, 
phenomenological and cosmological constraints. The direct search for the scattering 
of cold dark matter also excludes some regions of the NUHM parameter space, but 
the $\bmm$ constraint is stronger in most of the cases we have studied.

Experiments at the Tevatron and then the LHC are expected to increase greatly the
sensitivity to $\bmm$. Our analysis shows that, in many NUHM cases, this improved
sensitivity would have good prospects for detecting $\bmm$. For example, sensitivity
to $BR(\bmm) \sim 10^{-8}$ would give access to essentially all the $(\mu, M_A)$
planes for $m_{1/2} = 300$~GeV, $m_0 = 100$~GeV and $\tan \beta \ga 40$, as seen
in Fig.~\ref{fig:300_100}. On the other hand, in the case of larger $m_{1/2} = 500$~GeV and
$m_0 = 300$~GeV, shown in Fig.~\ref{fig:500_300}, the allowed regions of the WMAP strips
extend to larger values of $M_A$ that would require greater sensitivity to $BR(\bmm)$. The
same effect is seen even for the largest studied value of $\tan \beta = 57$ for several
different choices of larger values of $m_{1/2}$ and $m_0$, as seen in Fig.~\ref{fig:57}. 
Sensitivity to $BR(\bmm) \sim 10^{-8}$ would not be sufficient to explore any new region
of the $(M_A, \tan \beta)$ plane for the choice $\mu = 200$~GeV, $m_{1/2} = 250$~GeV and $m_0 = 1000$~GeV, but would explore all the allowed region for $\mu = 1000$~GeV, $m_{1/2} = 600$~GeV and $m_0 = 800$~GeV, as seen in panels (a) and (b) of Fig.~\ref{fig:mAtb}, respectively.

We conclude that $\bmm$ is already an important constraint on the NUHM parameter space,
and that, within this framework, it would have excellent prospects for a future detection of
indirect effects of supersymmetry.

\section*{Acknowledgments}
\noindent 
The work of K.A.O. and V.C.S. was supported in part
by DOE grant DE--FG02--94ER--40823. 
The work of Y.S. was supported in part
by the NSERC of Canada.

\end{document}